\documentclass[twocolumn,showpacs,superscriptaddress,amsmath,amssymb]{revtex4}

\usepackage{float,epsfig}

\begin{document}

\title{Self-Consistent Nuclear Shell-Model Calculation Starting from 
a Realistic $NN$ Potential}

\author{L. Coraggio}
\affiliation{Dipartimento di Scienze Fisiche, Universit\`a
di Napoli Federico II, \\ and Istituto Nazionale di Fisica Nucleare, \\
Complesso Universitario di Monte  S. Angelo, Via Cintia - I-80126 Napoli,
Italy}
\author{N. Itaco}
\affiliation{Dipartimento di Scienze Fisiche, Universit\`a
di Napoli Federico II, \\ and Istituto Nazionale di Fisica Nucleare, \\
Complesso Universitario di Monte  S. Angelo, Via Cintia - I-80126 Napoli,
Italy}
\date{\today}

\begin{abstract}
First self-consistent realistic shell-model calculation for the 
light $p$-shell nuclei is performed, starting from the
high-precision nucleon-nucleon ($NN$) CD-Bonn potential. 
This realistic potential is renormalized deriving a low-momentum $NN$ 
potential $V_{\rm low-k}$ that preserves exactly the two-nucleon 
low-energy physics. 
This $V_{\rm low-k}$ is suitable to derive a self-consistent
Hartree-Fock basis that is employed to derive both effective 
single-particle energies and residual two-body matrix elements 
for the shell-model hamiltonian.
Results obtained show the reliability of such a fundamental
microscopic approach.
\end{abstract}


\maketitle

The nuclear shell model is the foundation upon which the understanding of the
main features of the structure of the atomic nucleus is based.
In such a frame, a central role is performed by the auxiliary one-body
potential $U$, which has to be introduced in order to break up the
nuclear hamiltonian as the sum of a one-body component $H_0$, which
describes the independent motion of the nucleons, and a residual
interaction $H_1$:

\begin{equation}
H=\sum_{i=1}^{A} \frac{p_i^2}{2m} + \sum_{i<j} V_{ij} = T + V =
(T+U)+(V-U)= H_{0}+H_{1}~~.
\end{equation}

\noindent
Once $H_0$ has been introduced, it is possible to define a reduced
Hilbert space, the model space, in terms of a finite subset of $H_0$'s
eigenvectors. 
In this way the unfeasible task of diagonalizing the many-body
hamiltonian (1) in a infinite Hilbert space, may be reduced to the
one of solving an eigenvalue problem for an effective hamiltonian 
in a finite model space. 

During the last forty years, a lot of efforts have been devoted to 
derive effective shell-model hamiltonians, starting from free
nucleon-nucleon ($NN$) potentials $V_{NN}$, which reproduce with extreme
accuracy the $NN$ scattering data and the deuteron properties (see for
instance \cite{hjorth95}).
In particular, during the last ten years many realistic shell-model
calculations have been performed to describe quantitatively with a
great success the properties of nuclei over a wide mass range 
\cite{jiang92,engeland93,holt00,andreozzi97,coraggio99}.

In all these studies, for the sake of simplicity, a harmonic oscillator
potential has been adopted as auxiliary potential.
This choice simplifies the computation of the two-body interaction
matrix elements, as well as the derivation of the effective
hamiltonian, by way of a degenerate time-dependent pertubation theory
\cite{krenc80,suzuki80}. 

A more fundamental microscopic choice for $U$ is the Hartree-Fock (HF)
potential, that is self-consistently derived from the free potential 
$V_{NN}$.
Such a choice leads to a so-called self-consistent shell model 
\cite{zamick02}.
It is well known, however, that modern realistic $NN$ potentials
cannot be used directly to calculate a HF self-consistent
potential, owing to their strong repulsion at short distances.
A traditional approach to this problem is the Brueckner-Hartree-Fock (BHF)
procedure, where the self-consistent potential is defined in terms of the
reaction matrix vertices $G$ \cite{day67}.
However, this approach cannot be considered the basis for a fully 
self-consistent realistic shell-model calculation \cite{zamick02}, 
because the choice of the BHF potential for states above the Fermi 
surface cannot be uniquely defined \cite{towner}.

Recently, a new technique to renormalize the short-range behavior of 
a realistic $NN$ potential by integrating out its high-momentum
components has been introduced \cite{bogner02}. 
The resulting low-momentum potential $V_{\rm low-k}$ is a smooth 
potential that preserves the low-energy physics of $V_{NN}$ 
and can be used directly to derive a self-consistent HF potential 
\cite{coraggio03}. 
This paves the way to perform a full self-consistent realistic
shell-model calculation.

In this Letter, we present results of such a kind of shell-model
calculation for light $p$-shell nuclei, starting from the high 
precision CD-Bonn $NN$ potential \cite{cdbonn01}.

\begin{figure}[ht]
\includegraphics[scale=1.0,angle=90]{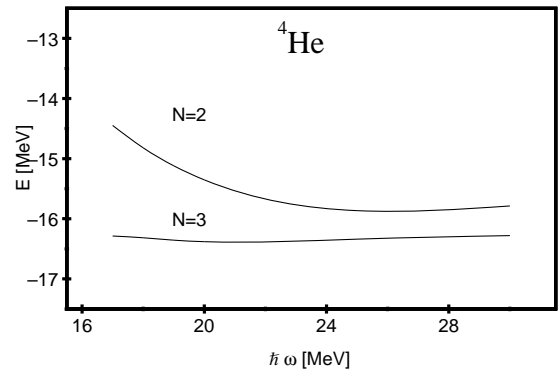}
\caption{Behavior of $E_{\rm HF}$ with $\hbar \omega$ and $N$ for 
$^{4}$He.}
\end{figure}

First, for the sake of clarity, we outline the path through which our
calculation winds up.
The first step consists in deriving from the CD-Bonn potential a
low-momentum potential $V_{\rm low-k}$ defined within a cutoff
momentum $\Lambda$. 
This is done in the spirit of the renormalization group theory so
obtaining a smooth potential which preserves exactly the on-shell
properties of the original $V_{NN}$ \cite{bogner02}. 
In Ref. \cite{bogner02} a detailed discussion has been done about 
the value of $\Lambda$ and a criterion for its choice.
According to this criterion, we have used here $\Lambda=2.1 \; 
{\rm fm}^{-1}$.
The so obtained $V_{\rm low-k}$, hermitized using the procedure based 
on Choleski decomposition suggested in Ref. \cite{andreozzi96}, has 
been used directly to solve the HF
equations for $^4$He doubly closed-shell core.
We remove the spurious center-of-mass kinetic energy \cite{davies71} 
writing the kinetic energy operator $T$ as

\begin{equation}
T= \frac{1}{2Am} \sum_{i<j} ({\rm {\bf p}}_i - {\rm {\bf p}}_j )^2 ~~.
\end{equation}

So, the hamiltonian can be re-written as

\begin{equation}
H= \left( 1 - \frac{1}{A} \right) \sum_{i=1}^{A} \frac{p_i^2}{2m} + \sum_{i<j}
\left( V_{ij} - \frac{ {\rm {\bf p}}_i \cdot {\rm {\bf p}}_j }{mA} \right) ~~.
\end{equation}

The details of our HF procedure may be found in Ref. \cite{coraggio03}.
In our calculations the HF single-particle (SP) states are expanded in
a finite series of $N=3$ harmonic-oscillator wave-functions for $^{4}$He.
This truncation is sufficient to ensure that the HF results do not 
significantly depend on the variation of the oscillator constant 
$\hbar \omega$, as shown in Fig. 1, where the behavior of the 
HF ground-state energy versus $\hbar \omega$ has been reported for 
different values of $N$. 
The value of $\hbar \omega$ adopted here is 18 MeV, as derived from
the expression $\hbar \omega= 45 A^{-1/3} -25 A^{-2/3}$ \cite{blomqvist68}.

\begin{table}[ht]
\caption{Experimental \cite{tilley02} and theoretical SP energies 
(referred to $^4$He closed-shell core) calculated including up to 
second- and third-order diagrams in $V_{\rm low-k}$. 
The experimental widths of the states are also given. Energies are in MeV.}
\begin{ruledtabular}
\begin{tabular}{ccccc}
 ~~ & 2nd order & 3rd order &   Expt.  & $\Gamma$ \\ 
\colrule
$\pi 0p_{3/2}$ &  1.109 &  1.355 & 1.97 & 1.23 \\
$\pi 0p_{1/2}$ &  4.800 &  4.970 & 3.46 & 6.60 \\
$\nu 0p_{3/2}$ & -0.081 &  0.202 & 0.89 & 0.65 \\
$\nu 0p_{1/2}$ &  3.796 &  3.984 & 2.16 & 5.57 \\
\end{tabular}
\end{ruledtabular}
\end{table}

The HF SP eigenvectors define the basis to be used for our shell-model
calculation.
More precisely, we assume that doubly-magic $^4$He is an inert core and
let the valence nucleons occupy the two HF orbits $0p3/2$ and $0p1/2$.

The next step is to derive the effective hamiltonian for the
chosen model space.
Starting from the time-dependent perturbation theory \cite{krenc80},
the effective hamiltonian is written in operator form as

\begin{equation}
H_{\rm eff} = \hat{Q} - \hat{Q'} \int \hat{Q} + \hat{Q'} \int \hat{Q} \int
\hat{Q} - \hat{Q'} \int \hat{Q} \int \hat{Q} \int \hat{Q} + ~...~~,
\end{equation}

\noindent
where $\hat{Q}$ is the irreducible vertex function $\hat{Q}$-box 
composed  of irreducible valence-linked diagrams, and the integral
sign represents a generalized folding operation.
$\hat{Q'}$ is obtained from $\hat{Q}$ by removing terms of first order 
in $V_{\rm low-k}$.

\begin{table}[ht]
\caption{Experimental and calculated energy levels, in MeV, in $^6$Li
and $^6$He. The calculated results
obtained including diagrams up to second and third order in 
$V_{\rm low-k}$ are compared. See text for comments.}
\begin{ruledtabular}
\begin{tabular}{cccc}
 $^{6}$Li & Expt. & 3rd order & 2nd order  \\ 
\colrule
Binding energy &  31.995 &  31.501 &  32.198 \\
$E_{\rm x}(1^+_1 0)$ & 0.0    & 0.0    & 0.0   \\
$E_{\rm x}(3^+ 0)$   & 2.186  & 1.912  & 1.757 \\ 
$E_{\rm x}(0^+ 1)$   & 3.563  & 2.627  & 2.401 \\
$E_{\rm x}(2^+ 0)$   & 4.312 & 5.215  & 5.170 \\
$E_{\rm x}(2^+_1 1)$ & 5.366  & 4.911  & 4.590 \\
$E_{\rm x}(1^+_2 0)$ & 5.65   & 7.825  & 7.305 \\
 & & & \\
 $^{6}$He & Expt. & 3rd order & 2nd order  \\ 
\colrule
Binding energy & 29.269  & 29.797 & 30.837 \\  
$E_{\rm x}(0^+_1 1)$ & 0.0  & 0.0    & 0.0   \\
$E_{\rm x}(2^+ 1)$   & 1.8  & 2.108  & 2.072 \\
\end{tabular}
\end{ruledtabular}
\end{table}

We take the $\hat{Q}$-box to be composed of one- and two-body diagrams
through third order in $V_{\rm low-k}$.

After the $\hat{Q}$-box is calculated, the energy-independent 
$H_{\rm eff}$ is then obtained by summing up the folded-diagram series 
of Eq. (4) to all orders using the Lee-Suzuki iteration method 
\cite{suzuki80}.
However, such a method is appropriate for degenerate model spaces
only, this is the case when using a harmonic oscillator basis, as 
pointed out before.

\begin{table}[ht]
\caption{Experimental and calculated energies (MeV) and electromagnetic 
properties in $^{7}$Li. The reduced transition probabilities are 
expressed in W.u..}
\begin{ruledtabular}
\begin{tabular}{ccc}
 $^{7}$Li & Expt. & Calc.   \\ 
\colrule
Binding energy &  39.243 &  39.556  \\
$E_{\rm x}(\frac{3}{2}^-_1 \frac{1}{2})$ & 0.0    & 0.0      \\
$E_{\rm x}(\frac{1}{2}^-_1 \frac{1}{2})$ & 0.478  & 0.498  \\
$E_{\rm x}(\frac{7}{2}^-_1 \frac{1}{2})$ & 4.652   & 3.998  \\ 
$E_{\rm x}(\frac{5}{2}^-_1 \frac{1}{2})$ & 6.604   & 5.993 \\
$E_{\rm x}(\frac{5}{2}^-_2 \frac{1}{2})$ & 7.454   & 7.213   \\
$E_{\rm x}(\frac{3}{2}^-_2 \frac{1}{2})$ & 8.75   & 8.985  \\
$E_{\rm x}(\frac{1}{2}^-_2 \frac{1}{2})$ & 9.09   & 9.899  \\
$E_{\rm x}(\frac{7}{2}^-_2 \frac{1}{2})$ & 9.57   & 9.564 \\
$E_{\rm x}(\frac{3}{2}^-_1 \frac{3}{2})$ & 11.24  & 9.206  \\
\colrule
$Q_{\rm gs}$ [$e$ mb]           & -40.6(8)&  -24.4  \\ 
$\mu_{\rm gs}$ [nm]    & +3.256  &  +4.28 \\
B(E2;$\frac{1}{2}^-_1  \rightarrow \frac{3}{2}^-_1$) & 19.7(1.2)  &  9.71   \\
B(E2;$\frac{7}{2}^-_1  \rightarrow \frac{3}{2}^-_1$) & 4.2  &  4.18   \\
B(M1;$\frac{1}{2}^-_1  \rightarrow \frac{3}{2}^-_1$) & 2.75(14)  &  2.37   \\
\end{tabular}
\end{ruledtabular}
\end{table}

The model space we are dealing with is non-degenerate, the unperturbed 
HF SP energies (respect to $^4$He closed-shell core) being 4.217 and 
7.500 MeV for $\pi 0p_{3/2}$ and $\pi 0p_{1/2}$, 3.106 and 6.577 MeV for 
$\nu 0p_{3/2}$ and $\nu 0p_{1/2}$, respectively.
So we use a generalization of the Lee-Suzuki iteration method to sum
up the folded diagram series, expressed, in the case of non-degenerate 
model spaces, in terms of the multi-energy $\hat{Q}$-boxes \cite{kuo95}. 
To our knowledge this is the first time that the above technique
has been employed to derive a realistic shell-model effective interaction.

As mentioned before, $H_{\rm eff}$ contains one-body contributions, the
sum of all these contributions (the so-called $\hat{S}$-box
\cite{shurpin83}) is what actually determines the SP
energies that should be used in a shell-model calculation.
It is customary, however, to use a subtraction procedure
\cite{shurpin83} so that only the two body terms of $H_{\rm eff}$ are
retained and the SP energies are taken from the experimental data
regarding the low lying spectra of the nuclei with one neutron or
proton outside the inert core.
In this calculation we have followed a more fundamental approach,
where the SP energies are the theoretical ones obtained from the 
$\hat{S}$-box calculation (see Table I).   

\begin{table}[ht]
\caption{Experimental and calculated electromagnetic properties in $^6$Li.
The reduced transition probabilities are expressed in W.u..}
\begin{ruledtabular}
\begin{tabular}{ccc}
 $^{6}$Li                           & Expt.      & Calc.   \\
\colrule
$Q_{\rm gs}$ [$e$ mb]               & -0.818(17) &  -0.442 \\
$\mu_{\rm gs}$ [nm]                 & +0.822     &  +0.866 \\
B(E2;$3^+_1 0 \rightarrow 1^+ 0$)   & 16.5(1.3)  &  6.61   \\
B(E2;$2^+ 0 \rightarrow 1^+_1 0$)   & 6.8(3.5)   & 6.45    \\
B(M1;$0^+ 1 \rightarrow 1^+_1 0$)   & 8.62(18)   & 9.01    \\
B(M1;$2^+_1 1 \rightarrow 1^+_1 0$) & 0.083(15)  &  0.154  \\
\end{tabular}
\end{ruledtabular}
\end{table}

\begin{table}[ht]
\caption{Same as Table IV, but for $^{7}$Be.}
\begin{ruledtabular}
\begin{tabular}{ccc}
 $^{7}$Be & Expt. & Calc.   \\ 
\colrule
Binding energy &  37.600 & 37.751  \\
$E_{\rm x}(\frac{3}{2}^-_1 \frac{1}{2})$ & 0.0      & 0.0       \\
$E_{\rm x}(\frac{1}{2}^-_1 \frac{1}{2})$ & 0.429    & 0.469  \\
$E_{\rm x}(\frac{7}{2}^-_1 \frac{1}{2})$ & 4.57(5)  & 3.922   \\
$E_{\rm x}(\frac{5}{2}^-_1 \frac{1}{2})$ & 6.73(10) & 5.782   \\
$E_{\rm x}(\frac{5}{2}^-_2 \frac{1}{2})$ & 7.21(6)  & 7.123   \\
\colrule
$\mu_{\rm gs}$ [nm]    & -1.398(15) & -1.013  \\
B(M1;$\frac{1}{2}^-_1  \rightarrow \frac{3}{2}^-_1$) & 2.07(27)  &  1.81   \\
\end{tabular}
\end{ruledtabular}
\end{table}

In Tables II-VIII we compare experimental binding energies, low-energy
spectra, and electromagnetic properties of $^6$Li, $^6$He, $^7$Li,
$^7$Be, $^8$Be, $^8$Li, and $^8$B \cite{selove88,audi93,tilley02} with 
calculated ones. 
All calculations have been performed using the OXBASH shell-model code 
\cite{oxb}. 
A quantitative amount of data for different nuclei is taken into
account in order to verify the reliability of a self-consistent
shell-model calculation.
As regards binding energy results, it is well known that shell-model 
calculations give ground state energies referred to the closed-shell
core.
In our approach, we can consistently calculate the $^4$He binding
energy by means of the Goldstone linked-cluster expansion \cite{coraggio03}.
So, our theoretical binding energies are obtained summing the ground
state energies of the open-shell nuclei to the calculated $^4$He
binding energy, whose value is 25.967 MeV including up to third-order 
contributions. 
Electromagnetic properties have been calculated using effective
operators \cite{krenc75} which take into account core-polarization effects.

\begin{table}[ht]
\caption{Experimental and calculated energies (MeV) in $^{8}$Be.}
\begin{ruledtabular}
\begin{tabular}{ccc}
 $^{8}$Be & Expt. & Calc.   \\ 
\colrule
Binding energy & 56.50 & 54.010 \\
$E_{\rm x}(0^+_1 0)$ & 0.0   & 0.0    \\
$E_{\rm x}(2^+_1 0)$ & 3.04  & 2.985  \\
$E_{\rm x}(4^+_1 0)$ & 11.40 & 10.056 \\
$E_{\rm x}(2^+_1 1)$ & 16.63 & 15.889 \\
$E_{\rm x}(2^+_2 0)$ & 16.92 & 15.174 \\
$E_{\rm x}(1^+_1 1)$ & 17.64 & 15.927 \\
$E_{\rm x}(1^+_1 0)$ & 18.15 & 15.070 \\
$E_{\rm x}(3^+_1 1)$ & 19.01 & 18.863 \\
$E_{\rm x}(3^+_1 0)$ & 19.24 & 16.983 \\
$E_{\rm x}(4^+_2 0)$ & 19.86 & 21.191 \\
\end{tabular}
\end{ruledtabular}
\end{table}

\begin{table}[ht]
\caption{Same as Table IV, but for $^{8}$Li.}
\begin{ruledtabular}
\begin{tabular}{ccc}
 $^{8}$Li & Expt. & Calc.   \\ 
\colrule
Binding energy & 41.276 & 44.365 \\
$E_{\rm x}(2^+_1 1)$ & 0.0   & 0.0        \\
$E_{\rm x}(1^+_1 1)$ & 0.981  & 0.922   \\
$E_{\rm x}(3^+_1 1)$ & 2.255 & 2.074 \\
$E_{\rm x}(1^+_2 1)$ & 3.21 & 4.667 \\
$E_{\rm x}(4^+_1 1)$ & 6.53 & 6.929  \\
$E_{\rm x}(0^+_1 2)$ & 10.822 & 8.230  \\
\colrule
$Q_{\rm gs}$ [$e$ mb]           & 24(2)&  26.7  \\ 
$\mu_{\rm gs}$ [nm]    & +1.653  &  +2.89 \\
B(M1;$1^+_1 1 \rightarrow 2^+_1 1$) & 2.8(9)  &  2.67   \\
B(M1;$3^+_1 1 \rightarrow 2^+_1 1$) & 0.29(13)  &  0.38   \\
\end{tabular}
\end{ruledtabular}
\end{table}

In Table II we compare calculated binding energies and spectra of $^6$Li and
$^6$He, obtained including contributions up to second- and third-order
in perturbation theory.
In all other tables calculated quantities refer to a third-order $H_{\rm eff}$.

From the inspection of Tables II-VIII, we can conclude that the
overall agreement between theory and experiment may be considered to
be quite satisfactory. 
This agreement is of the same quality, and in some cases even better, 
of that obtained in our previous work \cite{coraggio01}, where a realistic 
shell-model calculation for $p$-shell nuclei was carried out within the
framework of the semi-phenomenological two-frequency shell model.
It is worth to note that the inclusion of a realistic three-body force
could lead to an overall improvement in the agreement with experiment,
as shown in Refs. \cite{pieper01,navratil03}

\begin{table}[ht]
\caption{Experimental and calculated energies (MeV) and electromagnetic 
properties in $^{8}$B.}
\begin{ruledtabular}
\begin{tabular}{ccc}
 $^{8}$B & Expt. & Calc.   \\ 
\colrule
Binding energy & 37.74 & 40.429 \\
$E_{\rm x}(2^+_1 1)$ & 0.0   & 0.0        \\
$E_{\rm x}(3^+_1 1)$ & 2.32 & 2.061 \\
$E_{\rm x}(0^+_1 2)$ & 10.619 & 8.245  \\
\colrule
$Q_{\rm gs}$ [$e$ mb]           & 64.6(1.5)&  40.1  \\ 
$\mu_{\rm gs}$ [nm]    & +1.036  &  +1.73 \\
\end{tabular}
\end{ruledtabular}
\end{table}

This work was supported in part by the Italian Ministero
dell'Istruzione, dell'Universit\`a e della Ricerca  (MIUR).



\end{document}